%% file: sbdnn.tex
\documentclass[english,13pt]{article}
\usepackage[T1]{fontenc}
\usepackage[latin1]{inputenc}
\usepackage{geometry}
\usepackage{float}
\usepackage{mathrsfs}
\usepackage{amsmath}
\usepackage{bbm}
\usepackage{amssymb}
\usepackage{graphicx}
\usepackage{booktabs}
\usepackage{hyperref}
\hypersetup{
    colorlinks=true,
    linkcolor=blue,
    filecolor=magenta,      
    urlcolor=cyan,
    citecolor = red,
}

\makeatletter

\usepackage{algorithm,algpseudocode}

\usepackage{amsthm}

\usepackage{mathrsfs}

\usepackage{amsfonts}

\usepackage{epsfig}

\usepackage{bm}

\usepackage{mathrsfs}

\usepackage{enumerate}

\numberwithin{equation}{section}

\@ifundefined{definecolor}{\@ifundefined{definecolor}
 {\@ifundefined{definecolor}
 {\usepackage{color}}{}
}{}
}{}

\usepackage{subfig}\usepackage[all]{xy}

\newcounter{hypA}

\newcounter{hypB}

\newcounter{hypD}

\newcounter{hypW}

\usepackage{babel}\date{}

\usepackage{babel}

\makeatother

\usepackage{babel}

\newcommand{\sX}{\mathsf X}
\newcommand{\sY}{\mathsf Y}

\newcommand{\bbE}{\mathbb{E}}

\begin{document}

\begin{center}

{\Large \textbf{Score-Based Martingale Posteriors for Deep Neural Networks}}

\vspace{0.5cm}

BY  ABYLAY ZHUMEKENOV$^{1}$,  AJAY JASRA$^{1}$,  MOHAMED MAAMA$^{2}$ \& RAUL TEMPONE$^{2}$

{\footnotesize $^{1}$School of Data Science,  The Chinese University of Hong Kong,  Shenzhen,  Shenzhen, CN.}\\
{\footnotesize $^{2}$Applied Mathematics and Computational Science Program,  King Abdullah University of Science and Technology,  Thuwal,  KSA.}
{\footnotesize E-Mail:\,} \texttt{\emph{\footnotesize  abylayzhumekenov@cuhk.edu.cn
ajayjasra@cuhk.edu.cn; maama.mohamed@gmail.com; raul.tempone@kaust.edu.sa
}}

\end{center}

\begin{abstract}
In this paper we investigate the efficacy of the score-based martingale posteriors (SMP) \cite{cui2025martingale,fong2023martingale} in the context of modern and large-scale machine learning problems and its potential for meaningful uncertainty quantification. 
SMPs work with a stochastic gradient ascent-type recursion on the parameter space of stochastic models and construct a martingale on the parameter space.  Under simple mathematical assumptions,  the recursion can be built so that the parameters form a martingale sequence which possesses a limiting,  in time,  random variable,  the latter of which can be simulated very quickly,  in contrast to Monte Carlo-based methods such as Markov chain Monte Carlo.  In this expository paper we explore the SMP for inferring the parameters of deep neural networks (DNNs) and, where feasible, compare our results to the state-of-the-art Monte Carlo methods aimed at inferring conventional Bayesian posteriors.   
\end{abstract}

\section{Introduction}

Bayesian inference for unknown parameters of stochastic models is an important part of science,  with a huge number of applications such as machine learning,  medicine,  finance,  econometrics and many more;  see for example \cite{robert2007bayesian} for an introduction.   The standard Bayesian approach to statistical inference corresponds to placing a prior on the unknown parameters and updating the prior,  via the data likelihood resulting in the posterior density.  It is well-known that for many stochastic models of practical relevance,  analytic computations associated to the posterior are infeasible,  leading to a substantial literature on Monte Carlo-based computation methods,  such as Markov chain Monte Carlo and sequential Monte Carlo;  see for instance \cite{robert2004monte} and the references therein.  

In the context of modern applications,  with large data sets and sometimes quite complicated stochastic models,  whilst Monte Carlo methods have been continued to be developed (e.g.~diffusion-based samplers and Sinkhorn methods (see e.g.~\cite{del2026new,ruzayqat2023unbiased} and the references therein))  there appears a disconnect between the demands of applied researchers and what Monte Carlo methods can realistically provide.  Despite this,  the Bayesian idea is attractive,  in that it provides a coherent approach to uncertainty quantification for unknown parameters. 

One of the more interesting computationally fast methods for uncertainty quantification that has recently appeared is the martingale posterior idea \cite{cui2025martingale,fong2023martingale},  see also more recent work in \cite{fong2026asymptotics,moya2025martingale,ng2025tabmgp}.  In the guise of the method which we consider directly,  the idea is based upon stochastic gradient ascent-type recursion on the parameter space,  the score-based martingale posterior (SMP).  The idea is to run this recursion with the observed data, and for subsequent times (corresponding to the recursion) simulate data from the predictive distribution of the underlying stochastic model.  The recursion that is used is constructed so that one has,  ultimately,  a type of martingale and then,  using the well-known martingale convergence theorem,  that there exists a limiting (in time) random variable which one is ultimately simulating.  At least from a conceptual perspective,  this can be computationally cheaper than the afore-mentioned Monte Carlo methods.  

In this article the intention is to investigate SMP for the context of large-scale problems in machine learning.  Some investigation is contained in the work of \cite{lee2023martingale,ng2025tabmgp},  but the work in this article is more fundamental and focussed.  We consider deep neural network (DNN) models which in theory can (and have been \cite{chada2025bayesian}) fitted using Monte Carlo and develop and adapt the approach of \cite{cui2025martingale} for large scale inference.  The main idea is to understand whether,  fundamentally,  the method is a viable alternative to traditional Bayesian modeling at the small-scale and how useful the approach really is at the large scale.  

This paper is structured as follows.  In Section \ref{sec:mod} we outline the statistical model and the approach that is used for uncertainty quantification.  In Section \ref{sec:numerics}  we present several numerical results,  with a comparison,  in terms of computational time,  with state-of-the-art Bayesian computational methods for deep neural networks. Section~\ref{sec:disc} discusses practical insights and future research directions, and Section~\ref{sec:conc} concludes the paper.

\section{Modeling and Methodology}\label{sec:mod}

\subsection{Model}

We consider random variables of the form $\left((X_1,Y_1),(X_2,Y_2),\dots\right)$, where, for $i\in\mathbb{N}$, $X_i\in\sX\subseteq\mathbb{R}^n$ denotes a covariate and $Y_i\in\sY$ denotes a response. The objective is to infer a predictive model for the conditional distribution of $Y_i$ given $X_i$. We write this conditional model as
\begin{equation}
    f_\phi(y\mid x),
\end{equation}
where $\phi\in\Phi\subseteq\mathbb{R}^{d_\phi}$ is a collection of parameters. In our applications, $f_\phi(y\mid x)$ is induced by a deep neural network.

We also allow for a marginal model for the covariates. Let $f_\xi(x)$ denote a probability density or probability mass function on $\sX$, with parameter $\xi\in\Xi$. The space $\Xi$ may be a singleton, corresponding to the case where the distribution of the covariates is fixed and contains no unknown parameters. Writing
\begin{equation}
    \theta=(\phi,\xi),
\end{equation}
the joint model for a single pair $z=(y,x)$ is
\begin{equation}
    f_\theta(z) = f_\theta(y,x) = f_\phi(y\mid x)f_\xi(x).
\end{equation}
For a finite collection of $N$ observations, this gives
\begin{equation}
    f_\theta(y_{1:N},x_{1:N}) = \prod_{i=1}^N f_\phi(y_i\mid x_i)f_\xi(x_i), \qquad N\in\mathbb{N}.
\end{equation}
This formulation includes the usual fixed-design or empirical-design setting used in supervised learning. In that case, $f_\xi$ is fixed, there are no unknown covariate parameters, and the parameter vector reduces to $\theta=\phi$. This is the setting used in the numerical experiments of Section~\ref{sec:numerics}.

We now describe the conditional model $f_\phi(y\mid x)$. Suppose first that $\sY=\{0,\dots,m-1\}$, $m\in\mathbb{N}$. We let $g:\sX\times\Phi\rightarrow\mathbb{R}^m$ be a function with components $g(x,\phi)=(g_0(x,\phi),\dots,g_{m-1}(x,\phi))^\top$. The corresponding softmax probabilities are
\begin{equation}
    \pi_{\phi,j}(x) := \frac{\exp\{g_j(x,\phi)\}} {\sum_{k\in\sY} \exp\{g_k(x,\phi)\}}, \qquad j\in\sY.
\end{equation}
Then the categorical conditional likelihood is
\begin{equation}\label{eq:resp_class}
    f_\phi(y\mid x) = \pi_{\phi,y}(x), \qquad y\in\sY.
\end{equation}
For binary classification one may equivalently take $\sY=\{0,1\}$ and let $g:\sX\times\Phi\rightarrow\mathbb{R}$ be scalar-valued. Define
\begin{equation}
    p_\phi(x) := \sigma(g(x,\phi)), \qquad \sigma(u) = \frac{1}{1+\exp(-u)}.
\end{equation}
The conditional likelihood is then Bernoulli:
\begin{equation}\label{eq:resp_binary}
    f_\phi(y\mid x) = p_\phi(x)^y \{1-p_\phi(x)\}^{1-y}, \qquad y\in\sY.
\end{equation}
Regression models can also be handled in the same framework, for instance by taking $f_\phi(y\mid x)$ to be a Gaussian density with mean $g(x,\phi)$. In this paper, however, all numerical experiments are classification examples, and we therefore focus on the Bernoulli and categorical likelihoods above.

We now describe the construction of $g$ using a deep neural network (DNN); see, for example, \cite{goodfellow2016deep}. DNNs are constructed by composing affine functions with element-wise activation functions. Let $\nu:\mathbb{R}\rightarrow\mathbb{R}$ be an activation function and, for $k\in\mathbb{N}$, define $\zeta_k:\mathbb{R}^k\rightarrow\mathbb{R}^k$ by
\begin{equation}
    \zeta_k(z_1,\dots,z_k) = (\nu(z_1),\dots,\nu(z_k))^\top.
\end{equation}
For example, $\nu(u)=\max\{0,u\}$ gives the ReLU activation.

Let $D\in\mathbb{N}$ and let $(n_0,\dots,n_D)\in\mathbb{N}^{D+1}$ denote the layer sizes, with $n_0=n$ corresponding to the input dimension. Let $A_d\in\mathbb{R}^{n_d\times n_{d-1}}$ and $b_d\in\mathbb{R}^{n_d}$ denote the weights and biases of layer $d$, for $d\in\{1,\dots,D\}$. We use the notation $\phi := \left((A_1,b_1),\dots,(A_D,b_D)\right)$ and so $\phi \in \Phi = \bigotimes_{d=1}^D \left\{ \mathbb{R}^{n_d\times n_{d-1}}\times\mathbb{R}^{n_d} \right\}$. Define
\begin{equation}
    h_0(x,\phi)=x,
\end{equation}
and, for hidden layers $d=1,\dots,D-1$,
\begin{equation}
    h_d(x,\phi) = \zeta_{n_d}\left(A_d h_{d-1}(x,\phi)+b_d\right).
\end{equation}
The network output is then
\begin{equation}
    g(x,\phi) = A_D h_{D-1}(x,\phi)+b_D.
\end{equation}
For multiclass classification we take $n_D=m$, so that $g(x,\phi)\in\mathbb{R}^m$ gives the logits in the softmax model \eqref{eq:resp_class}. For binary classification one may take $n_D=1$, so that $g(x,\phi)$ is a scalar logit in \eqref{eq:resp_binary}.

Finally, we specify the measure notation used below. Let $\mu$ denote a dominating measure on $\sX\times\sY$. In continuous regression settings, $\mu$ may be Lebesgue measure on both components. In classification settings, $\mu$ is the product of a measure on $\sX$ and counting measure on $\sY$. In empirical-design settings, the measure on $\sX$ may be taken to be counting measure on the observed training covariates. Throughout, $f_\theta(z)$ denotes the density or probability mass function of $z=(y,x)$ with respect to this dominating measure. The collection of unknown parameters $\theta$ lies in the space $\Theta=\Phi\times\Xi$. We denote the dimension of $\Theta$ by $d_\theta$; in the fixed-design case used in the numerical experiments, $d_\theta=d_\phi$.

\subsection{Method}

\subsubsection{Underlying Recursion}

We will be using the methodology that is described in \cite{cui2025martingale}. The ideas are now given in the notation that has been established in the previous section. For $\theta\in\Theta$, let $F_{\theta}$ be the distribution on
$(\sX\times\sY,\mathscr{B}(\sX\times\sY))$ associated with the density or probability mass function $f_{\theta}(z)$ with respect to the dominating measure $\mu(dz)$. We assume that $f_{\theta}(z)$ is differentiable in $\theta$ for $\mu$-almost every $z\in\sX\times\sY$. Let $\theta_0\in\Theta$ be given and define the recursion for $k\in\mathbb{N}$
\begin{equation}\label{eq:sbm_rec}
\theta_k = \theta_{k-1} + \gamma_k \nabla \log f_{\theta_{k-1}}(Z_k),
\end{equation}
where for each $k\geq 1$,
\begin{equation}
Z_k \mid \theta_0,\theta_1,Z_1,\dots,\theta_{k-1},Z_{k-1}
\sim F_{\theta_{k-1}},
\end{equation}
with $\gamma_k\in\mathbb{R}^+$, $\sum_{k\geq 1}\gamma_k=\infty$, and $\sum_{k\geq 1}\gamma_k^2<\infty$. We remark that to compute the gradient vectors, one must use automatic differentiation as is standard in fitting DNN models.

In the fixed-design or empirical-design case used in the numerical experiments, the marginal distribution of the covariates contains no unknown parameters. Hence the parameter vector reduces to $\theta=\phi$, and the score in \eqref{eq:sbm_rec} is simply
\begin{equation}
\nabla_\theta \log f_{\theta}(Z_k)
=
\nabla_\phi \log f_{\phi}(Y_k\mid X_k),
\end{equation}
where $X_k$ is sampled from the fixed or empirical covariate distribution and $Y_k\sim f_{\phi}(\cdot\mid X_k)$.

We now justify the recursion \eqref{eq:sbm_rec} in the same manner as described in \cite{yao2026martingale}. Let $\mathbb{E}$ be the expectation operator associated to the afore-mentioned process $\theta_0,\theta_1,Z_1,\dots$. Let $\mathscr{F}_0$ be the trivial sigma field and, for $k\geq 1$, let $\mathscr{F}_k$ be the $\sigma$-field generated by $(\theta_0,\theta_1,Z_1,\dots,\theta_k,Z_k)$. Let $|\cdot|$ be the $L_1$-norm for vectors and suppose that:
\begin{itemize}
\item{$\sup_{k\geq 0}\mathbb{E}[|\theta_k|]<+\infty$;}
\item{$\mathbb{E}\left[\sup_{\theta\in\Theta}\left|\nabla \log f_{\theta}(Z_k)\right|\right] <+\infty.$}
\end{itemize}
Then, for any $k\geq 1$, one has
\begin{equation}
\mathbb{E}\left[\theta_k\mid\mathscr{F}_{k-1}\right]
=
\theta_{k-1},
\end{equation}
where we have assumed that it is legitimate to reverse the order of integration and differentiation and one has that $\int_{\sX\times\sY}\left\{\nabla \log f_{\theta}(z)\right\}f_{\theta}(z)\,\mu(dz)=0$ for each $\theta\in\Theta$. Therefore $(\theta_k)_{k\geq 0}$ is a martingale relative to $(\mathscr{F}_k)_{k\geq 0}$ and by, for example, \cite[Section 1.3]{hall1980martingale}, there exists an integrable random variable $\theta^{\star}$ such that almost surely
\begin{equation}
\lim_{k\rightarrow\infty}\theta_k = \theta^{\star}.
\end{equation}
In addition, suppose that for each $j\in\{1,\dots,d_{\theta}\}$:
\begin{itemize}
\item{$\mathbb{E}[(\theta_0^{(j)})^2]<+\infty$;}
\item{$\mathbb{E}\left[\sup_{\theta\in\Theta}\left(\bigl[\nabla \log f_{\theta}(Z_k)\bigr]^{(j)}\right)^2\right]< C$ for some $C<+\infty$.}
\end{itemize}
Then one can show that $(\theta^{\star})^{(j)}$ has finite variance.

Suppose that one runs the recursion \eqref{eq:sbm_rec} indefinitely. Then, under the stated assumptions, there is a limiting random variable which can be used to represent uncertainty about the model parameter. In practice one can sample \eqref{eq:sbm_rec} for a long time horizon and many times; the resulting collection of samples can be used as an approximation of $\theta^{\star}$. Further justifications are considered in \cite{fong2026asymptotics}.

\subsubsection{Initialising the Process, the Step-Size and Final Method}

We have not described how observed data can be integrated into the process.
In a similar manner to \cite{yao2026martingale} one could consider an initial training phase,  by running the recursion \eqref{eq:sbm_rec},  except one does not sample the data and uses the observed data in its place; this was adopted by \cite{cui2025martingale}. In this paper, however, we instead initialise from a high-quality point estimate. E.g. \cite{fong2026asymptotics} uses maximum likelihood estimates; for the toy example, we use the sample mean from a Hamiltonian Monte Carlo (HMC \cite{betancourt2017conceptual}) run, and for MNIST example, we use a maximum a posteriori (MAP) estimate. Then the martingale property only concerns the subsequent predictive-resampling phase.

In \cite{yao2026martingale} and the work of this article we have found that the step-size is an important factor associated to $\theta^{\star}$.   One method we consider is that of using a preconditioner (e.g.~\cite{li2017preconditioned,li2016preconditioned}),  see also \cite{fong2026asymptotics}:
\begin{equation}\label{eq:prec_update}
\theta_k = \theta_{k-1} + \gamma_k P_k^{-1}\nabla \log f_{\theta_{k-1}}(Z_k)    
\end{equation}
for some $d_{\theta}\times d_{\theta}$ matrix $P_k$, where $\gamma_k=\tau/(N+k)$ and $\tau>0$ is an additional scaling parameter. When available, one should use the Fisher information matrix of the model
\begin{equation}
    \mathcal{I}(\theta) = \bbE_{\theta}[s(\theta,Z)s(\theta,Z)^\top]
\end{equation}
where $s(\theta,Z)=\nabla\log f_{\theta}(Z)$ is the score. Otherwise, an approximation $P_k$ is needed. Note that if $P_k$ is $\mathscr{F}_{k-1}$-measurable, then
\begin{equation}
    \bbE\left[
    P_k^{-1}\nabla\log f_{\theta_{k-1}}(Z_k)\mid\mathscr{F}_{k-1}\right] 
    = P_k^{-1}\bbE\left[\nabla\log f_{\theta_{k-1}}(Z_k)\mid\mathscr{F}_{k-1}\right]
    = 0
\end{equation}
and we preserve the martingale property. We discuss the choices of preconditioner in Section~\ref{subsec:precond}. We present the final method for generating SMP samples in Algorithm \ref{alg:smp}.
\begin{algorithm}
\caption{Preconditioned SMP}\label{alg:smp}
\begin{algorithmic}
    \State \textbf{Input}: $\theta_0$, $P_1$, $\tau$, $S$, $K$, $N$
    \For{$\ell=1,\dots,S$} \textbf{in parallel}
        \For{$k=1,\dots,K$}
            \State Sample $X_k^{(\ell)}\sim f_{\xi_{k-1}^{(\ell)}}(\cdot)$
            \State Sample $Y_k^{(\ell)}\sim f_{\phi_{k-1}^{(\ell)}}(\cdot|X_k^{(\ell)})$
            \State Compute $s_k^{(\ell)}=\nabla\log f_{\theta_{k-1}^{(\ell)}}(Z_k^{(\ell)})$ where $Z_k^{(\ell)}=(Y_k^{(\ell)},X_k^{(\ell)})$
            \State Update $\theta_k^{(\ell)}=\theta_{k-1}^{(\ell)}+\frac{\tau}{N+k} (P_k^{(\ell)})^{-1} s_k^{(\ell)}$
            \State Update $P_{k+1}^{(\ell)}$
        \EndFor
    \EndFor
    \State \Return $\{\theta_K^{(1)},\dots,\theta_K^{(S)}\}$
\end{algorithmic}
\end{algorithm}

\section{Numerical Results}\label{sec:numerics}

We now assess the behaviour of SMP for neural-network classifiers in two regimes. The first is a low-dimensional synthetic binary classification problem, where NUTS (No-U-Turn Sampler, an adaptive implementation of HMC, \cite{hoffman2014no}) provides a feasible Bayesian reference. The second is MNIST, where full Bayesian sampling is computationally prohibitive and SMP is compared against a strong MAP baseline. The experiments are intended to evaluate both predictive performance and the quality of uncertainty quantification.

\subsection{Experimental setting}\label{subsec:exp_setting}

All numerical experiments use the empirical-design version of the framework in Section~\ref{sec:mod}. Since no marginal covariate parameter is used in these experiments, the full parameter vector coincides with the neural-network parameter vector. Thus, throughout Section~\ref{sec:numerics}, we write $\theta=\phi$ and use $f_\theta(y\mid x)$, $g(x,\theta)$, $p_\theta(x)$ and $\pi_{\theta,j}(x)$ for the corresponding conditional likelihood, logits, binary success probability and multiclass probabilities.

During the SMP predictive-resampling phase, covariates are sampled with replacement from $N=N_{\mathrm{train}}$ observed training inputs. Equivalently, with respect to counting measure on the observed training covariates, we use
\begin{equation}
    \widehat f_{\mathrm{emp}}(x) = \frac{1}{N_{\mathrm{train}}}\sum_{i=1}^{N_{\mathrm{train}}}\delta_{x_i}(x).
\end{equation}
Thus, at martingale step $k$,
\begin{equation}
    X_k \sim \widehat f_{\mathrm{emp}},
    \qquad
    Y_k \mid X_k,\theta_{k-1}\sim f_{\theta_{k-1}}(\cdot\mid X_k),
\end{equation}
and the score in Algorithm~\ref{alg:smp} becomes
\begin{equation}
    s_k = \nabla \log f_{\theta_{k-1}}(Y_k\mid X_k).
\end{equation}

For each SMP configuration, the recursion is initialised at a point estimate $\theta_0$ specified separately for each example. In the present fixed-design setting, this is equivalently a neural-network parameter estimate $\theta_0=\phi_0$. We then run only the predictive-resampling phase; no observed-data phase is used. Each SMP sample is obtained by truncating the recursion after the experiment-specific number of martingale steps.

\subsubsection{Preconditioners}\label{subsec:precond}
We distinguish between the \emph{structure} of the preconditioner (which determines which entries are non-zero) and the \emph{temporal strategy} (how the numerical values are obtained). Structures:
\begin{itemize}
\item None: $P_k = I$ (unpreconditioned).
\item Diagonal: $P_k = \operatorname{diag}(\widehat{\mathcal{I}}_k)$, where $\widehat{\mathcal{I}}_k$ is some approximation of the Fisher information matrix at iteration $k$.
\item Block-diagonal: $P_k = \operatorname{blockdiag}(\widehat{\mathcal{I}}_k)$, with blocks corresponding to either weight or bias parameters of each layer.
\item Dense: $P_k = \widehat{\mathcal{I}}_k$ (full matrix).
\end{itemize}
Temporal strategies:
\begin{itemize}
\item Fixed: compute $\widehat{\mathcal{I}}_k = \widehat{\mathcal{I}}(\theta_0)$ once using the full training set and keep it constant for all $k>1$.
\item EMA (exponential moving average): update online as $\widehat{\mathcal{I}}_k = \beta\,\widehat{\mathcal{I}}_{k-1} + (1-\beta) \Delta_k$, where $\Delta_k=s_{k-1}s_{k-1}^\top$ is the Fisher contribution from $(X_{k-1},Y_{k-1})$, and $\beta=0.98$.
\item Periodic: recompute $\widehat{\mathcal{I}}_k$ from scratch every $T$ iterations using the current parameter value $\theta_{k-1}$ and the full training set.
\end{itemize}
A small ridge $\lambda>0$ is added to the diagonal of all preconditioners for numerical stability; specific values are given per experiment.

\subsubsection{Predictive metrics}\label{subsec:metrics}

For each sampling-based method we obtain parameter samples $\{\theta^{(\ell)}\}_{\ell=1}^S$. For NUTS these are Bayesian posterior samples, whereas for SMP they are independent realisations of the truncated martingale recursion. For point-estimate baselines, the predictive distribution is computed from a single parameter value.  
At a test point $x$, the predictive distribution (averaged over the samples) is
\begin{equation}
\widehat{f}_\theta(y\mid x) = \frac{1}{S}\sum_{\ell=1}^S f_{\theta^{(\ell)}}(y\mid x),
\end{equation}
which approximates the Bayesian posterior predictive distribution (or, for SMP, the martingale posterior predictive).  
We evaluate the quality of these predictions using several complementary metrics, each capturing a different aspect of predictive performance and uncertainty quantification. These are: accuracy, Brier score, negative log-likelihood (NLL), expected calibration error (ECE), predictive entropy ($H_{\mathrm{pred}}$), expected entropy ($H_{\mathrm{exp}}$), mutual information (MI), and epistemic standard deviation. All metrics are computed over the test set $\{(x_i, y_i)\}_{i=1}^{N_{\mathrm{test}}}$. Detailed definitions are provided in Appendix~\ref{app:metrics}.

By reporting this set of metrics, we can disentangle the sources of uncertainty: accuracy and proper scoring rules assess point predictions, calibration measures reliability, and entropy-based quantities separate aleatoric from epistemic uncertainty. This allows a fair comparison between standard Bayes (NUTS), the point-estimate MAP, and the SMP variants.

\subsubsection{Implementation}

The code was written in Julia language (v1.12.4), using \texttt{Turing.jl} (v0.45.0) for running NUTS, \texttt{Lux.jl} (v1.31.4) for neural network and SMP implementations, and \texttt{Zygote.jl} (v0.7.10) as the automatic differentiation backend. The source code is available upon request. All experiments were run on a Fedora 43 workstation with 6 Intel i7-8700 CPUs and 16 GB of RAM. We set the number of Julia threads to 6 for SMP sampling in both examples, and to 1 for NUTS in the toy example. The reported timings are for single-threaded NUTS and multi-threaded SMP. In addition, each parallel chain gets its own unique random seed.

\subsection{Toy example: 2D binary classification}\label{subsec:toy_example}

\subsubsection{Data generation and network architecture}

We generate $N_{\mathrm{train}}=500$ training points by drawing covariates $X_i$ uniformly from $[-1,1]^2$. The corresponding binary labels $Y_i$ are generated as
\begin{equation}
Y_i \sim \text{Bernoulli}\left(p_{\mathrm{true}}(X_i)\right),\qquad p_{\mathrm{true}}(x) = \sigma\left(20\times(0.5 - \|x\|_2^2)\right),
\end{equation}
where $\sigma(\cdot)$ is the logistic function (sigmoid). This creates a circular decision boundary with high confidence inside the circle ($\|x\|_2^2<0.5$) and outside ($\|x\|_2^2>0.5$).

The neural network follows the architecture described in Section~\ref{sec:mod} with $D=3$ layers, input dimension $n=2$, output dimension $m=1$, and layer sizes $(n_0,n_1,n_2,n_3) = (2,3,2,1)$. We use the GELU activation function (Gaussian error linear unit, \cite{hendrycks2016gaussian}) on the first two hidden layers; the output layer has no activation (it produces the logit $g(x,\theta)$). The total number of parameters is $d_\theta = 20$. The conditional likelihood is Bernoulli:
\begin{equation}
f_\theta(y\mid x) = p_\theta(x)^y\{1-p_\theta(x)\}^{1-y}, \qquad p_\theta(x)=\sigma(g(x,\theta)).
\end{equation}

\subsubsection{Baselines: NUTS and point estimate}

We compare SMP against two baselines: a full Bayesian posterior approximation using NUTS and a point estimate derived from it.

For the Bayesian posterior, we run NUTS with a single chain. The prior is $\mathcal{N}_{d_\theta}(0,I)$ (independent standard Gaussians). We initialise the chain using default \texttt{Turing.jl} settings, perform $1000$ warm-up (adaptation) steps, and then draw $S = 1000$ samples. The target acceptance rate is set to $0.95$. The NUTS diagnostics showed no numerical errors or divergent transitions. The acceptance rate was close to the target value, and the tree depth remained below the default maximum, giving no indication of numerical instability. Posterior summaries are computed from these samples after discarding the warm-up, both parameter and logit effective sample sizes are close to $1000$. We note that with only one chain, the NUTS samples characterise a local posterior mode and should not be interpreted as the true posterior distribution, which may be multimodal. This approach is intentional, as our goal is a local posterior approximation within one basin, aligning with the local nature of SMP.

To obtain a point estimate that lies in the same basin as the NUTS samples (for fair comparison with SMP), we take the posterior mean
\begin{equation}
\widehat{\theta} = \frac{1}{S}\sum_{\ell=1}^{S} \theta^{(\ell)}.
\end{equation}
We initialise all SMP runs with $\theta_0 = \widehat{\theta}$, and we also report the predictive performance of this point estimate itself (by using it as a single sample). In tables and plots we refer to this baseline as \texttt{NUTS-mean}. This ensures that differences between SMP and NUTS are not due to starting from different regions of the parameter space.

\begin{figure}[ht]
    \centering
    \includegraphics[width=\linewidth]{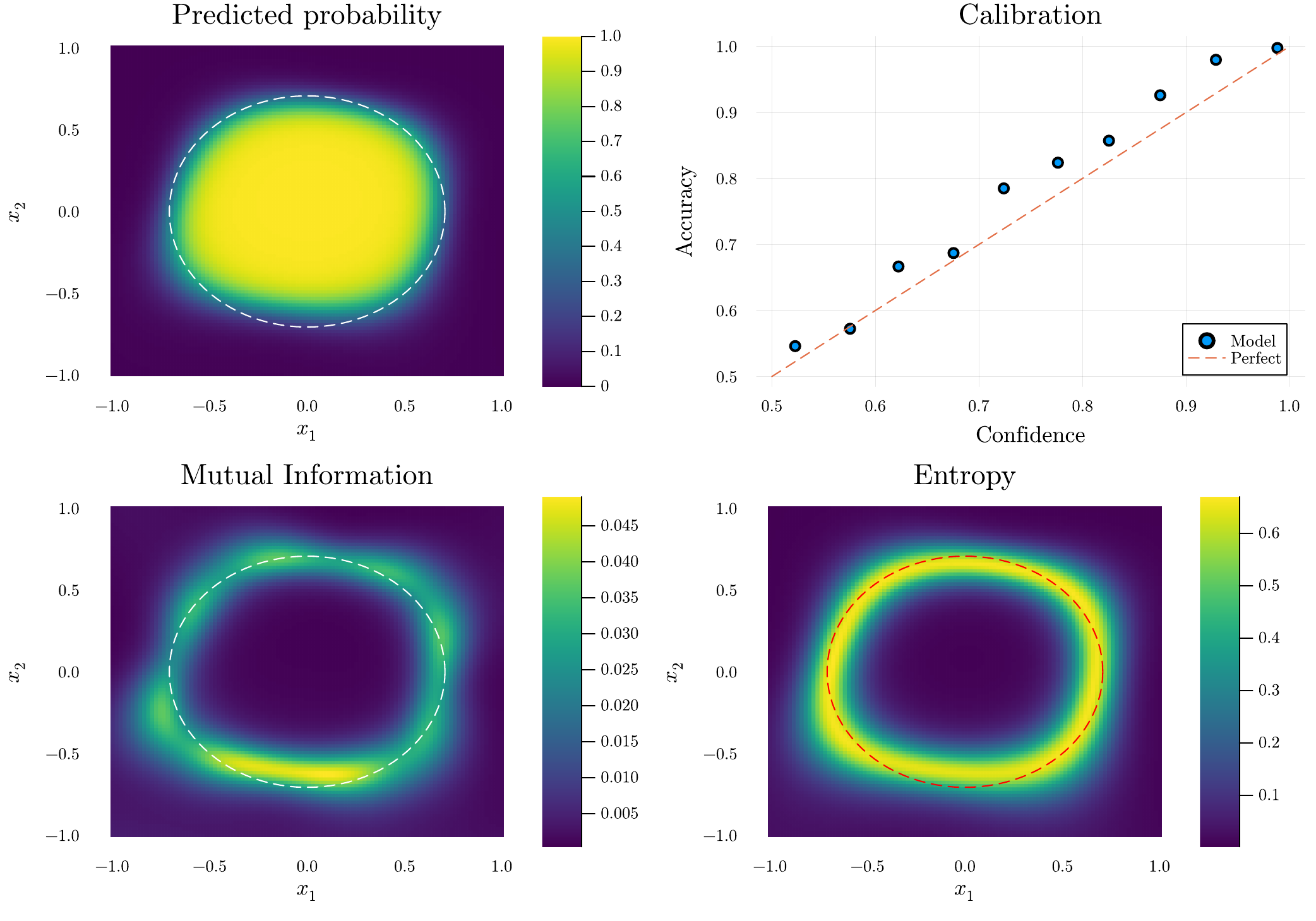}
    \caption{(Top-left) Predicted probabilities, (top-right) calibration plot, (bottom-left) mutual information, (bottom-right) expected entropy. \texttt{NUTS}. Toy example.}
    \label{fig:ex1_hmc_pred}
\end{figure}

In Figure~\ref{fig:ex1_hmc_pred}, we present predicted probabilities, mutual information and expected entropy on a grid of test points, as well as a calibration plot. These function-space summaries are useful because the neural-network parameter space is difficult to interpret directly due to symmetries, non-identifiability, and possible multimodality; predictive summaries instead describe the induced classifier and its uncertainty. The predicted probability surface almost correctly recovers the circular structure of the data-generating mechanism, with probabilities close to one inside the circle, close to zero outside it, and a smooth transition between the two classes around the boundary $\|x\|_2^2=0.5$. The expected entropy is largest near this boundary, where the true class probability is close to 0.5, reflecting high aleatoric uncertainty in the data-generating process. The mutual information is also concentrated near the boundary, indicating that posterior samples disagree most about the location and shape of the decision surface. However, its magnitude is much smaller than that of the expected entropy, suggesting that the dominant source of predictive uncertainty is the intrinsic label noise near the boundary, rather than epistemic uncertainty about the model parameters. Finally, the calibration plot, computed using $J=10$ confidence bins and $N_{\mathrm{test}}=5000$ new test points, lies mostly close to the diagonal reference line, indicating that the predictive probabilities are reasonably calibrated, although some bin-to-bin variation remains.

\begin{figure}[ht]
    \centering
    \includegraphics[width=\linewidth]{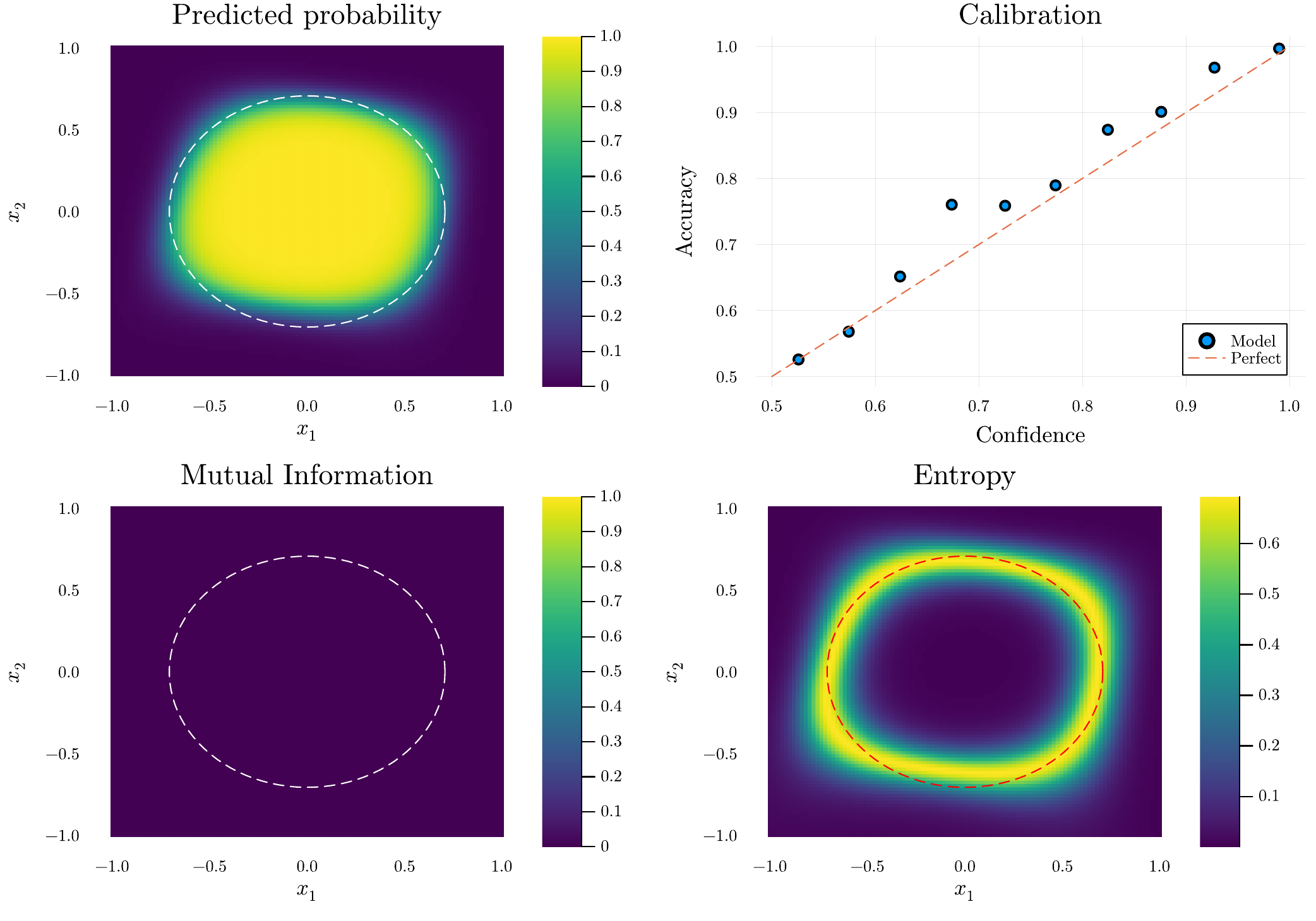}
    \caption{(Top-left) Predicted probabilities, (top-right) calibration plot, (bottom-left) mutual information, (bottom-right) expected entropy. \texttt{NUTS-mean}. Toy example.}
    \label{fig:ex1_map_pred}
\end{figure}

In Figure~\ref{fig:ex1_map_pred}, we have similar plots for the point estimate. The results are almost identical, except there is zero mutual information and the network is slightly more confident.

\subsubsection{SMP implementation details}

For the SMP experiments, we initialise all runs with $\theta_0 = \widehat{\theta}$, and we generate $S = 1000$ independent samples. Each sample is obtained by running the recursion for $K = 5000$ iterations (all simulated, no observed-data phase). We can define $r_K^2=\sum_{i>K}(N_{\mathrm{train}}+i)^{-2}$ and compute $r_K^2/r_0^2$ as a rough proxy for the residual variance ratio, but the quantity does not include the effects of $\tau$, the preconditioner, or the score covariance. In our case $r_K^2/r_0^2\times100\%\approx9\%$.

The step-size scaling factor $\tau$ is set to $\{1,0.3,0.3,0.1\}$ for the none, diagonal, block-diagonal and dense preconditioners, respectively (chosen based on a preliminary sweep to avoid numerical instability). A small ridge $\lambda = 10^{-4}$ is added to the diagonal of all preconditioners. The reason why we introduce these modifications is that the true Fisher and its approximation are often ill-conditioned and might lead to unstable trajectories.

We implement all combinations of the three preconditioner structures (none, diagonal, block, dense) and three temporal strategies (fixed, EMA, periodic) as described in Section~\ref{subsec:precond}. For the EMA strategy we set $\beta = 0.98$; for periodic we recompute every $T = 500$ iterations. The Fisher information estimate $\widehat{\mathcal{I}}_k$ is computed using the model-expected Fisher (averaged over the full training set for fixed and periodic, or online for EMA). For the binary classification with logistic likelihood, the model-expected Fisher information matrix conditional on covariates is
\begin{equation}
\widehat{\mathcal{I}}(\theta) = \frac{1}{N_{\mathrm{train}}} \sum_{i=1}^{N_{\mathrm{train}}} p_\theta(x_i)(1-p_\theta(x_i)) \left(\nabla g(x_i,\theta)\right)\left(\nabla g(x_i,\theta)\right)^\top.
\end{equation}

\subsubsection{Main results}

We evaluate all SMP variants (four preconditioner structures: none, diagonal, block-diagonal, dense; and three temporal strategies: fixed, EMA, periodic) on the toy binary classification problem, and compute the predictive metrics described in Section~\ref{subsec:metrics}. The results are summarised in Tables~\ref{tab:ex1_scores}, \ref{tab:ex1_variances} and \ref{tab:ex1_elapsed}.

\input{tables/ex1_scores.tex}
\input{tables/ex1_variances.tex}
\input{tables/ex1_elapsed.tex}

Among all SMP variants, the combination of a block-diagonal preconditioner with the EMA (exponential moving average) temporal strategy -- denoted \texttt{SMP-ema-block} provides the best compromise between predictive performance and non-degenerate epistemic uncertainty. Its test accuracy (0.9292) is nearly identical to \texttt{NUTS-mean} (0.9304) and \texttt{NUTS} (0.9304). The Brier score (0.0524) and NLL (0.1867) are only slightly worse than those of \texttt{NUTS} (0.0501 and 0.1672, respectively). The ECE (0.0451) is about twice that of \texttt{NUTS} (0.0210), indicating a modest but noticeable miscalibration. In contrast, fixed and periodic diagonal preconditioners produce accuracy close to \texttt{NUTS} but with negligible mutual information (MI $<0.015$), meaning they collapse to an almost deterministic posterior. The dense preconditioner, even with a reduced step-size ($\tau=0.1$), exhibits high ECE (0.1698 for fixed, 0.0925 for periodic) and inflated MI (0.3847 and 0.2230), suggesting over-dispersion and poor calibration.

The mutual information of \texttt{SMP-ema-block} (0.0458) is about four times larger than that of \texttt{NUTS} (0.0115) and substantially higher than \texttt{NUTS-mean} (0.0). The expected entropy ($H_{\mathrm{exp}}$) -- which captures average aleatoric uncertainty -- is 0.2403 for \texttt{SMP-ema-block}, moderately higher than \texttt{NUTS} (0.2034) and \texttt{NUTS-mean} (0.2006). This indicates that \texttt{SMP-ema-block} attributes a larger share of predictive uncertainty to parameter (epistemic) variation. The epistemic standard deviation of the logits for \texttt{SMP-ema-block} is 0.0973, roughly 2.5 times that of \texttt{NUTS} (0.0388). These numbers suggest that while \texttt{SMP-ema-block} produces meaningful epistemic uncertainty, it tends to be somewhat over-dispersed relative to the Bayesian reference.

Table~\ref{tab:ex1_variances} reports the variance of the posterior samples in parameter space (averaged over dimensions) and in logit space (averaged over test points). \texttt{SMP-ema-block} has a parameter variance of 0.1768, less than half of \texttt{NUTS} (0.3985). Nevertheless, its logit variance (1.6986) is about 1.8 times larger than \texttt{NUTS} (0.9526). This discrepancy reflects the non-linear mapping from parameters to predictions: small parameter changes near flat directions can cause disproportionately large predictive changes. Consequently, certain SMP variants can produce meaningful epistemic uncertainty even with comparatively compact parameter samples.

Table~\ref{tab:ex1_elapsed} reports wall-clock times. \texttt{NUTS} required 405.0 seconds to obtain 1000 samples using a single chain. The fastest SMP variant (no preconditioner) finished in 107.4 seconds when run with six threads. \texttt{SMP-ema-block} took 163.4 seconds -- a speed-up of about $2.5$ relative to \texttt{NUTS}. Even the most expensive variant (periodic dense, 286.6 seconds) can run faster than \texttt{NUTS} when multi-threading. In terms of total CPU time the comparison is less favorable, but SMP is embarrassingly parallel, better suited to parallel hardware and can achieve better wall-clock timings.

\begin{figure}[ht]
    \centering
    \includegraphics[width=\linewidth]{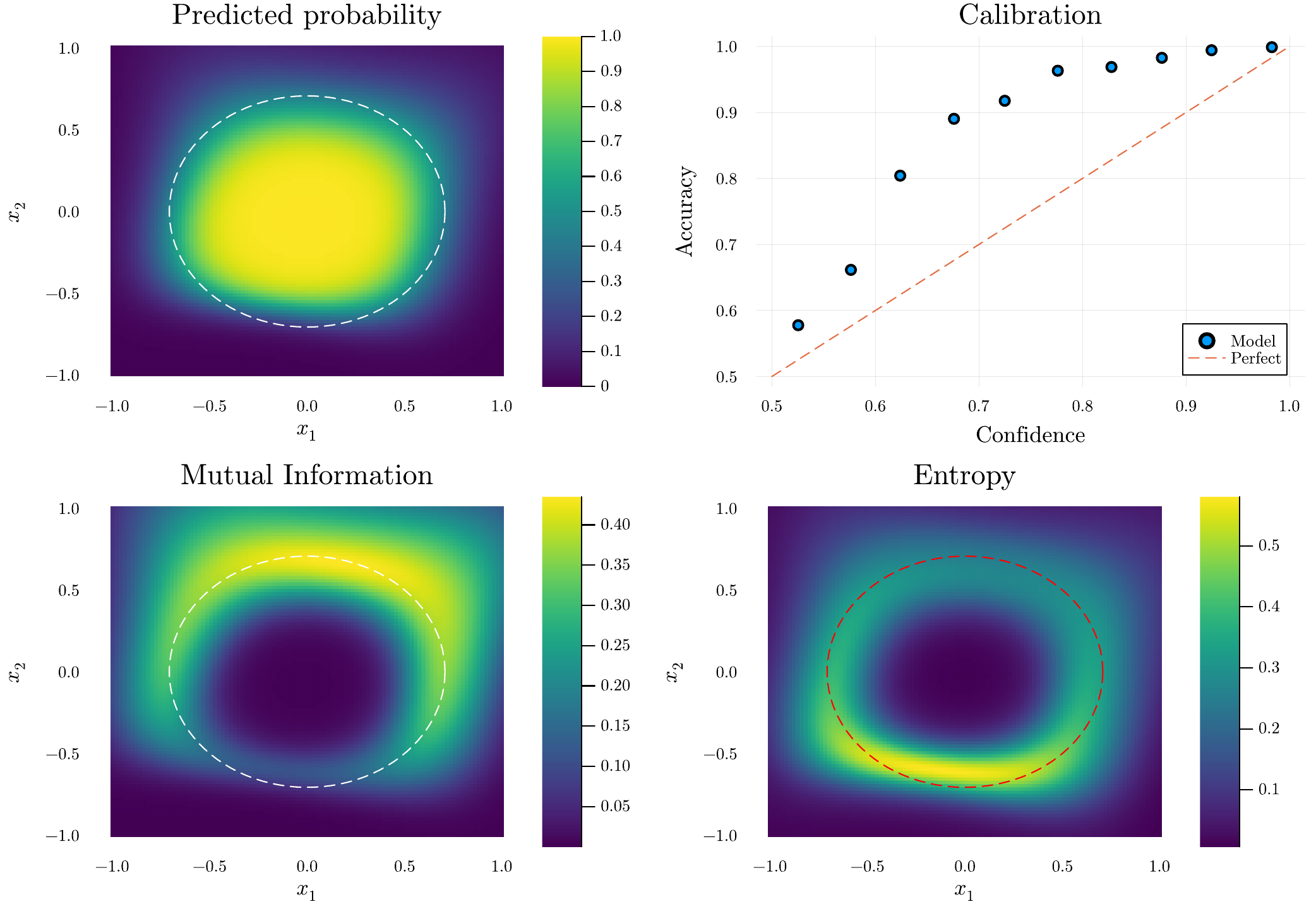}
    \caption{(Top-left) Predicted probabilities, (top-right) calibration plot, (bottom-left) mutual information, (bottom-right) expected entropy. \texttt{SMP-none}. Toy example.}
    \label{fig:ex1_smp0_pred}
\end{figure}

\begin{figure}[ht]
    \centering
    \includegraphics[width=\linewidth]{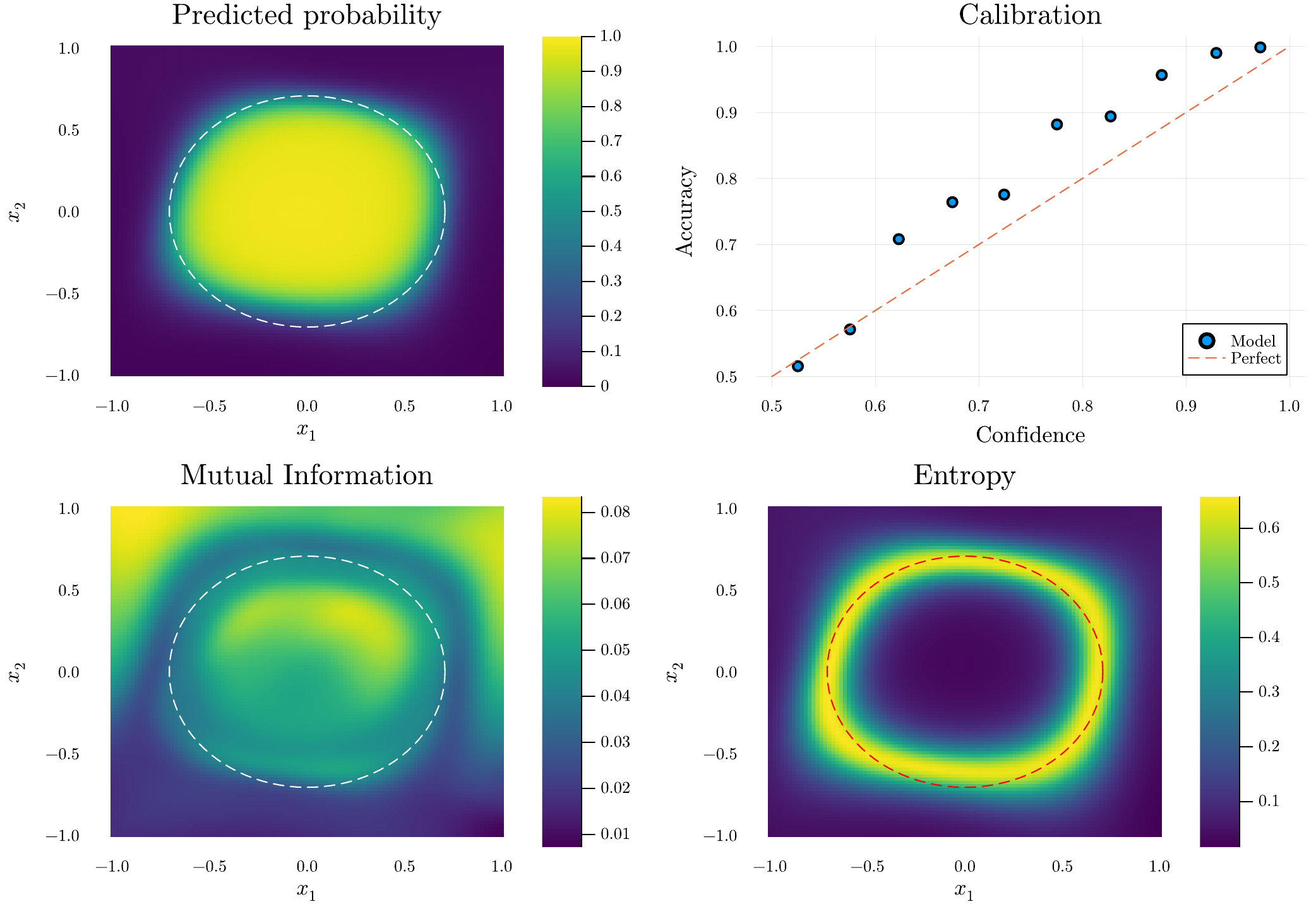}
    \caption{(Top-left) Predicted probabilities, (top-right) calibration plot, (bottom-left) mutual information, (bottom-right) expected entropy. \texttt{SMP-ema-block}. Toy example.}
    \label{fig:ex1_smp5_pred}
\end{figure}

To showcase the importance of the preconditioner, we present predictions made by \texttt{SMP-none} and \texttt{SMP-ema-block} in Figures \ref{fig:ex1_smp0_pred} and \ref{fig:ex1_smp5_pred}, respectively. The predicted probabilities of \texttt{SMP-none} are much more diffuse and the decision boundary is smoothed out. This does not affect accuracy, as the location of the boundary did not change much, but the network is less confident than it needs to be, indicating a miscalibration issue. We can identify the source of this excessive uncertainty by looking at mutual information and expected entropy plots. The entropy looks smoother, but has a similar magnitude to that of \texttt{NUTS}. On the other hand, the maximum MI value is around 10 times larger (0.45 vs 0.045). This means that the extra uncertainty is mostly epistemic rather than aleatoric, i.e. coming from the disagreement between samples about predictions. One reason might be that there is much more variance in predicted logits compared to \texttt{NUTS} (9.4562 vs 0.9526). The variance of parameters is much lower (0.0022 vs 0.3985), meaning that \texttt{SMP-none} collapses to the point estimate $\theta_0$, but movements in certain parameter directions swing the predictions dramatically.

On the other hand, \texttt{SMP-ema-block} is better informed about the geometry of the posterior. It makes predictions that can closely match those of \texttt{NUTS}; probabilities and entropy are visually identical. The calibration plot is slightly worse, but is adequate overall. Mutual information plot has an interesting feature -- uncertainty is actually lower right on the boundary than elsewhere. A possible interpretation might be that the samples agree on the location of the decision boundary, but there is some variation of predictions deep inside the classes. This explains why the sampler can explore the function space without hurting the accuracy.

\begin{figure}[ht]
    \centering
    \includegraphics[width=0.75\linewidth]{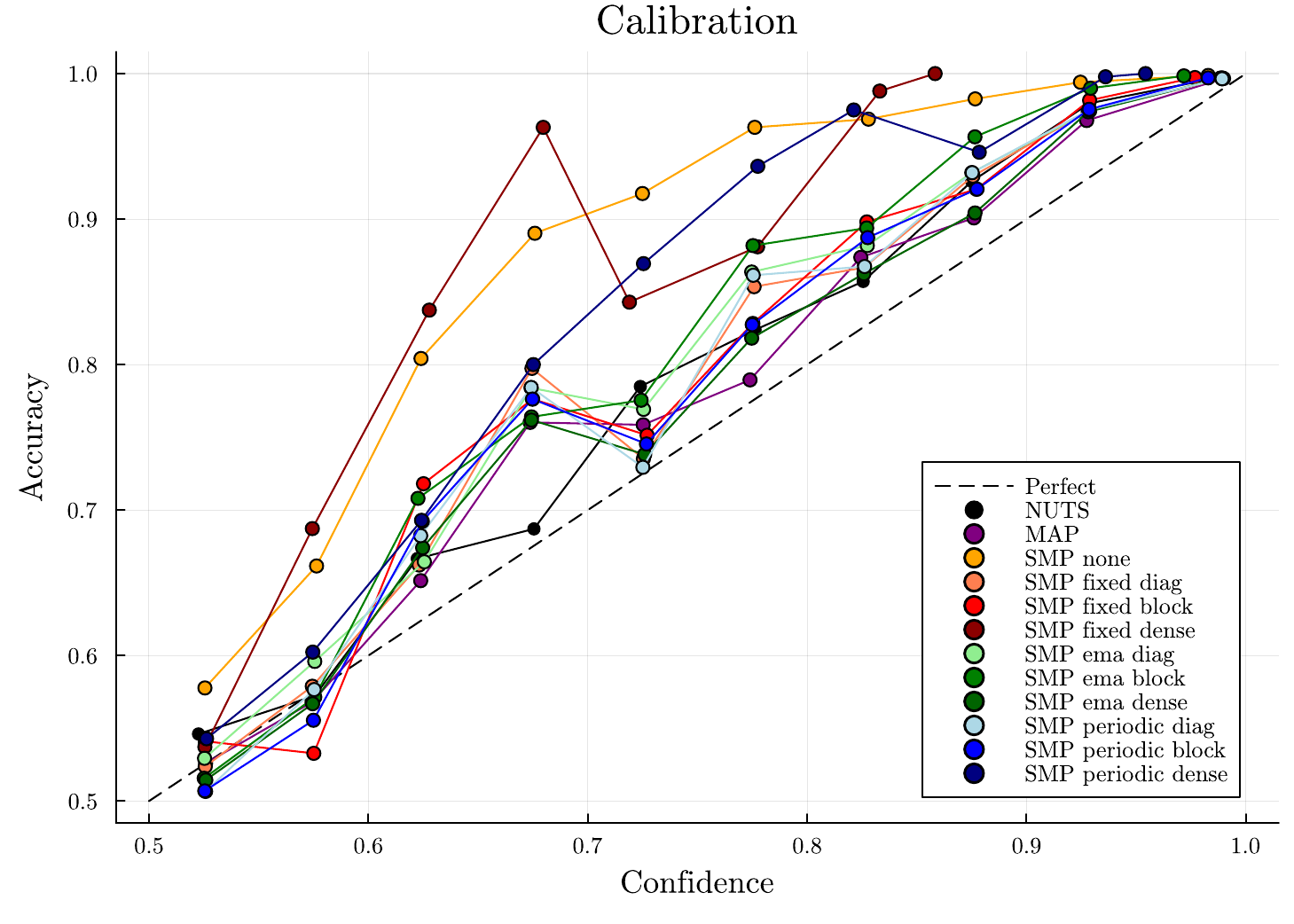}
    \caption{Calibration plots for all methods. Toy example.}
    \label{fig:ex1_calibration}
\end{figure}

Figure \ref{fig:ex1_calibration} compares the calibration of all methods. Unsurprisingly, \texttt{NUTS} and \texttt{NUTS-mean} are among the best calibrated, while \texttt{SMP-none} and \texttt{SMP-fixed-dense} show the largest deviation. \texttt{SMP-ema-block} lies close to \texttt{NUTS}, consistent with Table \ref{tab:ex1_scores}. Based on these results, we select \texttt{SMP-ema-block} as the recommended configuration for the toy example. It delivers predictive accuracy comparable to the Bayesian local posterior while providing a non-degenerate quantification of epistemic uncertainty, at a computational cost that is lower than HMC and allows easy parallelisation.

\subsection{MNIST example}\label{subsec:mnist_example}

\subsubsection{Data and network architecture}

The MNIST dataset consists of $28 \times 28$ grayscale images of handwritten digits $0$-$9$, split into $N_{\mathrm{train}}=54{,}000$ training, $N_{\mathrm{val}}=6{,}000$ validation, and $N_{\mathrm{test}}=10{,}000$ test samples. Inputs are scaled to $[0,1]$ and treated as vectors in $\sX = \mathbb{R}^{784}$. The output space is $\sY=\{0,\dots,9\}$, corresponding to digit labels.

We adopt a convolutional architecture similar to LeNet-5 \cite{lecun1998gradient}. The network consists of two convolutional layers (each with $5 \times 5$ kernels, ReLU activation, and $2 \times 2$ max-pooling) which output a flattened vector of size $256$, followed by two fully connected layers of sizes $128$ and $84$ with ReLU activations, and a final linear layer to $10$ logits. The total number of parameters is $d_\theta = 47{,}154$, with most of the parameters in later layers. The conditional likelihood is categorical:
\begin{equation}
f_\theta(y \mid x) = \pi_{\theta,y}(x), \qquad y \in \{0,\dots,9\}.
\end{equation}
with $\pi_{\theta,y}$ being a softmax probability.

\subsubsection{Baseline: MAP estimate}

We first obtain a point estimate $\theta_0=\widehat{\theta}$ by minimising the negative log-likelihood with $L_2$ regularisation (weight decay $10^{-3}$). Training uses the Adam optimiser with batch size $48$ and initial learning rate $10^{-3}$ for $20$ epochs, at which point the validation loss starts to grow again. This MAP-type point estimate achieves $98.77\%$ test accuracy, a Brier score of $0.0189$, and an NLL of $0.0380$. Its ECE is negligible ($0.0012$), and as expected it provides no epistemic uncertainty (mutual information and epistemic standard deviation are zero, see Table \ref{tab:ex2_scores}). This serves as a strong point-prediction baseline and is used to initialise all SMP runs.

\subsubsection{Feasible SMP variants}

For a network of this size, the dense and block-diagonal preconditioners are infeasible. Storing a $47{,}154 \times 47{,}154$ Fisher matrix would require approximately $9$ GB of memory (in single precision) before accounting for factorisations or additional workspace, and solving linear systems with it is computationally prohibitive. Even the block version has more than $30{,}000$ parameters in its largest layer, and it would be equally infeasible. We therefore restrict our attention to the unpreconditioned variant ($P_k = I$) and the diagonal preconditioner ($P_k = \operatorname{diag}(\widehat{\mathcal{I}}(\theta))$). For the diagonal case we consider all three temporal strategies: fixed, EMA, and periodic.

The model-expected Fisher for this model would require us to additionally compute Jacobians of the network, which is expensive. Thus, we use the empirical Fisher as an approximation. Computing the full empirical Fisher over the entire training set at each periodic recomputation is also prohibitively expensive. We therefore approximate its diagonal using a fixed set of $N_{\mathrm{val}} = 6{,}000$ held-out validation examples, but representing the same distribution as the training data. This yields a computationally feasible estimate of the curvature while preserving the same data distribution for the preconditioner. For the fixed and periodic strategies, the diagonal of the empirical Fisher is computed as
\begin{equation}
\operatorname{diag}\bigl(\widehat{\mathcal{I}}_{\mathrm{emp}}(\theta)\bigr) = \frac{1}{N_{\mathrm{val}}} \sum_{i=1}^{N_{\mathrm{val}}} \bigl( \nabla_\theta \log f_\theta(y_i \mid x_i) \bigr)^{\odot 2},
\end{equation}
where $\odot 2$ denotes the element-wise square. For the EMA strategy, we update the preconditioner as described in Section~\ref{subsec:precond}, using simulated $(X_k,Y_k)$. 

We set the number of martingale steps to $K = 10^{6}$. The step-size scaling factor is $\tau\in\{1,0.05,0.08,0.05\}$ for the unpreconditioned, fixed, EMA and periodically updated diagonal variants, all with $\lambda=10^{-4}$ regularisation. We generate $S = 100$ independent posterior samples using six parallel threads.

\subsubsection{Results}

Table~\ref{tab:ex2_scores} reports the predictive metrics for the MAP baseline and the four SMP variants (unpreconditioned, fixed diagonal, EMA diagonal, periodic diagonal). Table~\ref{tab:ex2_variances} shows the mean parameter variance and logit variance (averaged over test images). Table~\ref{tab:ex2_elapsed} gives the wall-clock times for generating posterior samples with six parallel threads.

\input{tables/ex2_scores.tex}
\input{tables/ex2_variances.tex}
\input{tables/ex2_elapsed.tex}

\texttt{SMP-none} attains the same accuracy ($98.77\%$) and almost identical Brier ($0.0190$) and NLL ($0.0380$) as \texttt{MAP}. Its calibration is similarly excellent (ECE $0.0011$). The mutual information ($0.0002$) and epistemic standard deviation ($0.0004$) are tiny, indicating that the martingale posterior almost collapses to a point estimate. The parameter variance is effectively zero ($\approx 6\times10^{-6}$) and the logit variance is only $0.0087$. This variant is computationally expensive ($4130$ seconds) but produces practically the same predictions as \texttt{MAP}.

All three diagonal variants show substantially degraded predictive performance compared to \texttt{MAP} and \texttt{SMP-none}, despite using carefully tuned step-size scaling factors ($\tau=0.05$ for fixed/periodic, $\tau=0.08$ for EMA). 
\begin{itemize}
    \item \texttt{SMP-fixed-diag}: Accuracy drops to $98.37\%$, Brier skyrockets to $0.4478$, NLL to $0.9907$, and ECE to $0.6050$. The MI ($1.877$) and logit variance ($101.9$) are enormous, indicating an extremely over-dispersed posterior that assigns probability almost uniformly across classes for many test points.
    \item \texttt{SMP-ema-diag}: Accuracy is only $97.97\%$, Brier $0.1808$, NLL $0.4582$, ECE $0.3266$. MI ($0.9079$) and logit variance ($18.88$) remain very high, though lower than the fixed variant.
    \item \texttt{SMP-periodic-diag}: This performs best among the diagonal variants ($98.63\%$ accuracy, Brier $0.0312$, NLL $0.0940$, ECE $0.0581$). However, MI ($0.1737$) and logit variance ($2.78$) are still orders of magnitude larger than those of \texttt{SMP-none} or \texttt{MAP}.
\end{itemize}
All three diagonal variants exhibit parameter variances of order $10^{-4}$ or smaller, yet their logit variances are huge. This suggests that the diagonal Fisher preconditioner, even when regularised and applied with a reduced step-size, induces movements in parameter directions that catastrophically amplify uncertainty in the prediction space. The effect is most severe for the fixed and EMA strategies, while periodic recomputation every $500$ iterations partially mitigates the issue but does not eliminate it.

Table~\ref{tab:ex2_elapsed} shows that \texttt{SMP-none} required $4130$ seconds (about $1.15$ hours) to generate $100$ posterior samples with six threads. The diagonal variants are even slower ($4620-6902$ seconds) due to the additional overhead of computing the diagonal Fisher (periodic recomputation is the most expensive). In contrast, the MAP point estimate required only training time (not reported) and no posterior sampling. For comparison, a full Bayesian inference with NUTS on this network would be prohibitively expensive and is therefore omitted.

\begin{figure}[ht]
    \centering
    \includegraphics[width=0.75\linewidth]{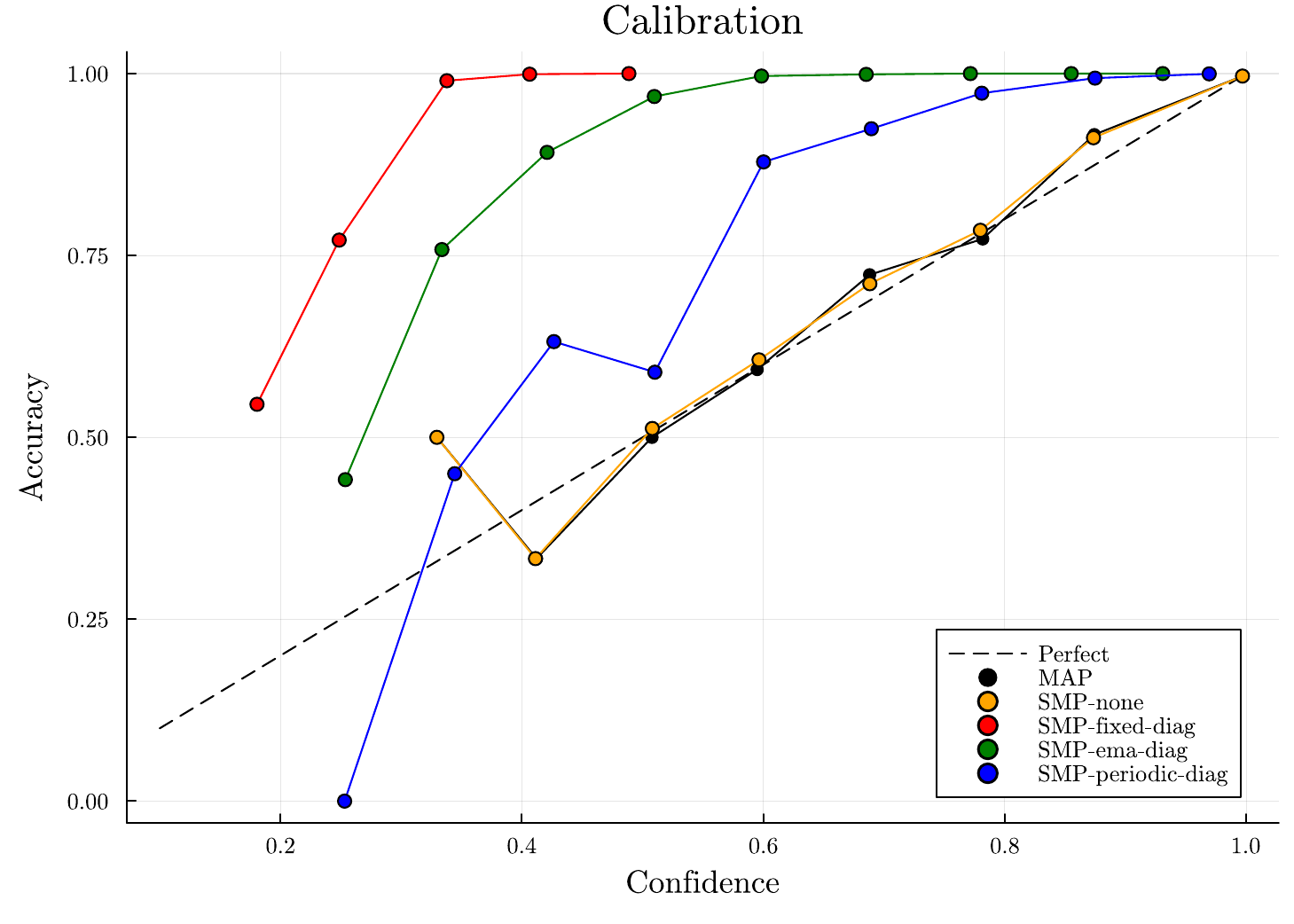}
    \caption{Calibration plots for MAP and all SMP variants. MNIST example.}
    \label{fig:ex2_calibration}
\end{figure}

Figure~\ref{fig:ex2_calibration} compares the calibration curves of all five methods. \texttt{MAP} and \texttt{SMP-none} lie almost perfectly on the diagonal, confirming their excellent calibration. The diagonal variants show severe miscalibration, with the fixed and EMA strategies producing predicted confidences that are far lower than the empirical accuracy (points far above the diagonal), while the periodic variant lies in between.

\begin{figure}[ht]
    \centering
    \includegraphics[width=\linewidth]{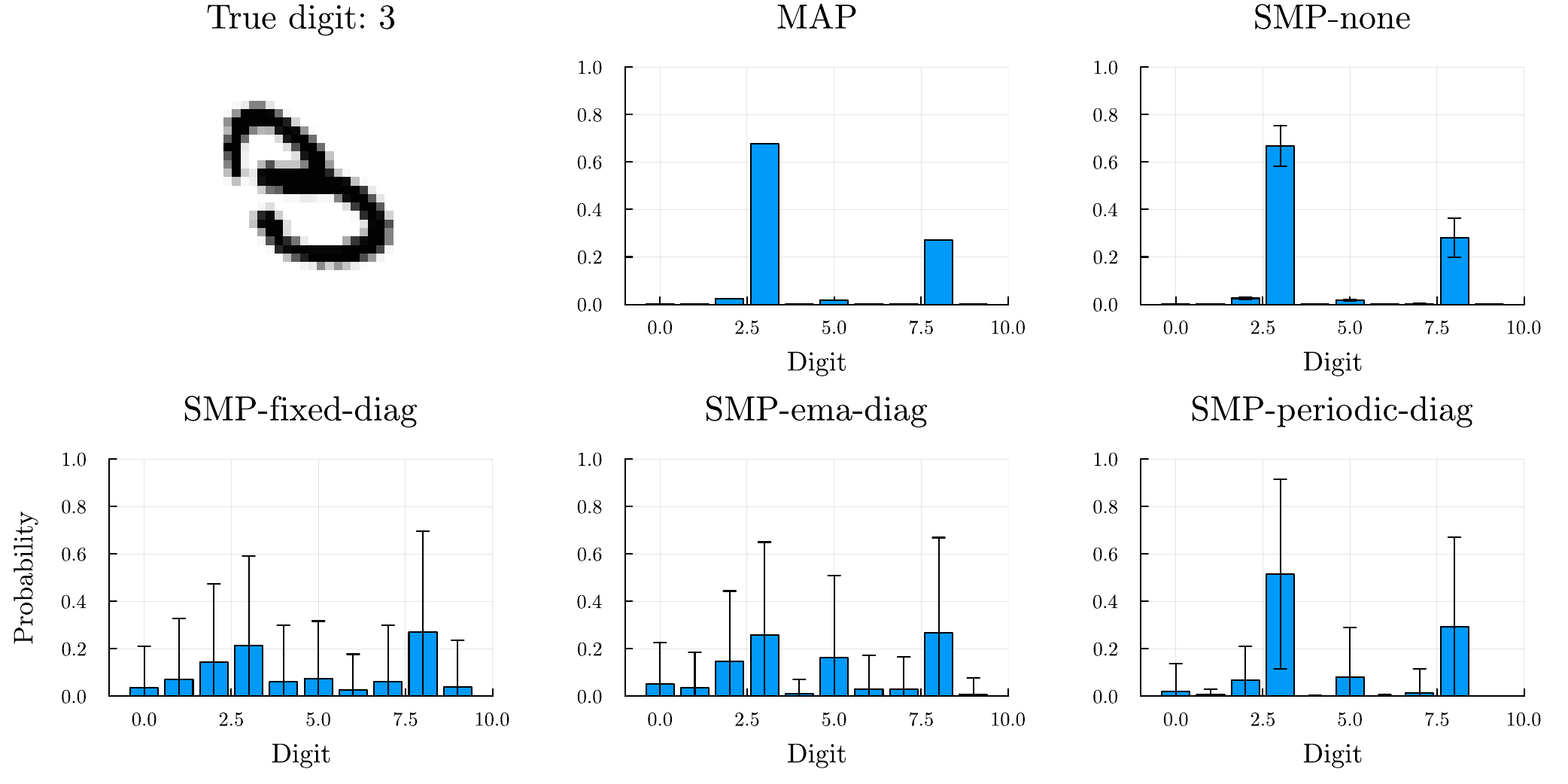}
    \caption{Predicted class probabilities for a test image with ambiguous label (true label ``3''). MNIST example.}
    \label{fig:ex2_predictions}
\end{figure}

Figure~\ref{fig:ex2_predictions} shows the predicted probabilities for a selected test image (a digit ``3'') across methods. \texttt{SMP-none} produces a sharp distribution concentrated on the correct class, similar to \texttt{MAP}. In contrast, the diagonal variants produce diffuse probability vectors, with \texttt{SMP-fixed-diag} being closer to uniform, \texttt{SMP-ema-diag} moderately diffuse, and \texttt{SMP-periodic-diag} slightly less so but still far from confident.

For the large-scale MNIST task, the only reliable SMP variant is the unpreconditioned one, which yields predictions nearly identical to the MAP estimate at the cost of a modest (though not negligible) computational overhead. Diagonal preconditioners, regardless of temporal strategy, lead to severe over-dispersion and poor calibration. This contrasts sharply with the toy example, where the block-diagonal EMA preconditioner succeeded. We hypothesise that the diagonal approximation of the Fisher information matrix is too crude for deep convolutional networks: it ignores cross-parameter correlations that are essential for stable curvature scaling, causing the martingale recursion to step into directions that massively inflate predictive uncertainty. Periodic recomputation helps but does not fully solve the problem. For practical uncertainty quantification on large neural networks, the unpreconditioned SMP provides a safe, computationally feasible baseline, but more sophisticated preconditioners (e.g., KFAC or layer-wise block approximations) would be needed to obtain meaningful epistemic uncertainty without collapsing or exploding.

\section{Discussion}\label{sec:disc}

In this paper, we investigated the score-based martingale posterior (SMP) for deep neural networks, comparing its performance to Bayesian MCMC (NUTS) and point estimates (NUTS mean and MAP) on a toy binary classification problem and the MNIST dataset. The two experiments reveal distinct regimes where SMP can succeed or struggle, and highlight practical guidelines for its application to deep neural networks.

\subsection{Preconditioner choice}
In the small toy example (20 parameters), the block-diagonal preconditioner with an EMA estimate of the Fisher produced meaningful epistemic uncertainty and accuracy nearly matching NUTS. The dense preconditioner was unstable, while the diagonal variant collapsed to a point estimate. For MNIST ($47{,}154$ parameters), block and dense preconditioners are infeasible; diagonal variants (fixed, EMA, periodic) all led to severe over-dispersion and miscalibration, whereas unpreconditioned SMP reproduced the MAP point estimate accurately but offered no uncertainty quantification. This suggests that a crude diagonal Fisher approximation is insufficient for deep convolutional architectures; layer-wise block approximations (e.g., KFAC) may be necessary for stable uncertainty quantification.

\subsection{Step-size tuning}
The step-size $\tau$ critically controls the variance of the limiting martingale posterior. For the toy example, $\tau = 0.3$ (diagonal/block) and $\tau = 0.1$ (dense) balanced stability and exploration. For MNIST, unpreconditioned SMP tolerated $\tau = 1$, where diagonal preconditioners required much smaller values ($0.05$--$0.08$) to avoid blow-up. A grid search over $\tau$ is recommended for new architectures. While it is possible for the toy model, for MNIST a full grid is prohibitive (we used a coarse manual sweep guided by stability).

\subsection{Activation functions}
The toy example used GELU, which breaks scaling symmetries and improved Fisher conditioning. MNIST used ReLU (standard for convnets). The poor performance of diagonal preconditioners on MNIST may be partly due to ReLU's scaling symmetries exacerbating ill-conditioning. A closer look at values on the diagonal of the estimated Fisher revealed that more than half of them were actual zeros. For preconditioned SMP, we would recommend smooth, non-scale-invariant activations (GELU, Swish) that do not easily saturate, but more future work is needed to give a definitive answer.

\subsection{Computational cost and practical trade-offs}
SMP samples are embarrassingly parallel. In the toy example, \texttt{SMP-ema-block} was $2.5\times$ faster than a single-chain NUTS run (163 vs 405 seconds, six vs one thread). Unpreconditioned MNIST SMP required 1.15 hours for 100 samples (six threads) -- still potentially faster than a full Bayesian treatment. However, several challenges arise:
\begin{itemize}
    \item GELU is more expensive than ReLU (Gaussian CDF evaluation), though acceptable for small nets.
    \item Memory constraints: a full Fisher matrix for MNIST would require $\approx 9$ GB; even diagonal preconditioners need careful memory management for gradient storage over the validation set.
    \item Tuning $\tau$ and generating samples for large nets remains non-trivial; future work should explore GPU acceleration and mini-batch approximations.
    \item Number of required SMP iterations before truncation scales roughly proportionally to the training data size, which makes sampling challenging where the data is abundant.
\end{itemize}
The hybrid sampling scheme (predictive resampling with a Gaussian tail approximation) is a promising direction from \cite{fong2026asymptotics}, but we did not implement it in either the toy or MNIST experiments. In our work, we used only basic truncation of the martingale recursion. The hybrid scheme would likely reduce the required number of martingale steps (e.g., from $10^6$ to a few thousand) and thus the runtime, without compromising the asymptotic distribution, if the Fisher preconditioner captures the local geometry well. We leave its integration for future work.

\subsection{Limitations}

We conclude the discussion by acknowledging two limitations of our study.

First, our comparison focused on exact Bayesian inference (NUTS) where feasible, and on a strong point estimate (MAP) otherwise. We did not benchmark SMP against other popular approximate uncertainty methods for deep neural networks, such as Monte Carlo Dropout, deep ensembles, or Laplace approximations. These methods are computationally lighter than MCMC and are known to provide practical uncertainty estimates. A systematic comparison between SMP and such baselines would be valuable to position SMP within the broader landscape of scalable uncertainty quantification, but it is beyond the scope of this work.

Second, unlike the Bayesian posterior obtained via NUTS, the SMP does not incorporate an explicit prior distribution. The limiting distribution of the martingale recursion is determined solely by the likelihood and the step-size schedule, which can be viewed as an implicit prior induced by the algorithm. Consequently, our comparison between SMP and the Bayesian posterior is not perfectly aligned: differences in uncertainty estimates (e.g., overdispersion or collapse) could partly arise from the absence of prior regularization rather than from the SMP recursion itself. For applications where prior information is available, extending SMP to incorporate a prior is an interesting direction for future research.

\section{Conclusion}\label{sec:conc}

SMP can be a fast, parallelisable alternative to MCMC for Bayesian uncertainty quantification in neural networks, but its success is highly sensitive to the preconditioner. For large convolutional networks, unpreconditioned SMP is safe but yields no epistemic uncertainty; obtaining meaningful uncertainty likely requires more sophisticated curvature approximations (e.g., KFAC) or a different architecture (e.g., using normalisation layers, smooth activations, or smaller layer sizes). Our results provide practical guidance for practitioners and highlight open challenges for future research.

\pagebreak
\bibliographystyle{abbrv}
\bibliography{references}

\pagebreak
\appendix

\section{Predictive metrics}\label{app:metrics}

\begin{itemize}
\item Accuracy (higher is better):
\begin{equation}
\mathrm{Acc} = \frac{1}{N_{\mathrm{test}}}\sum_{i=1}^{N_{\mathrm{test}}} \mathbbm{1}\Bigl( \arg\max_{j\in\sY} \widehat{f}_\theta(j\mid x_i) = y_i \Bigr).
\end{equation}
This measures raw classification performance but ignores confidence or uncertainty.

\item Brier score (lower is better):
\begin{equation}
\mathrm{Brier} = \frac{1}{N_{\mathrm{test}}}\sum_{i=1}^{N_{\mathrm{test}}}\sum_{j\in\sY} \bigl( \widehat{f}_\theta(j\mid x_i) - \mathbbm{1}(y_i=j) \bigr)^2.
\end{equation}
A proper scoring rule that penalises overconfidence: it is minimized when the predicted probabilities match the true outcome probabilities.

\item Negative log-likelihood (lower is better):
\begin{equation}
\mathrm{NLL} = -\frac{1}{N_{\mathrm{test}}}\sum_{i=1}^{N_{\mathrm{test}}}\log \widehat{f}_\theta(y_i\mid x_i).
\end{equation}
Another proper scoring rule that heavily penalises over-confident wrong predictions. Lower values indicate better fit.

\item Expected calibration error (lower is better): group predictions into $J$ bins according to the predicted confidence $\max_{j\in\sY} \widehat{f}_\theta(j\mid x)$. Let $B_j$ be the set of indices in bin $j$. Then
\begin{equation}
\mathrm{ECE} = \sum_{j=1}^{J}\frac{|B_j|}{N_{\mathrm{test}}}\bigl|\mathrm{acc}(B_j)-\mathrm{conf}(B_j)\bigr|,
\end{equation}
where $\mathrm{acc}(B_j)$ is the fraction of correct predictions in bin $j$ and $\mathrm{conf}(B_j)$ is the average confidence in that bin. ECE measures whether predicted probabilities are reliable (e.g., a 70\% confidence should correspond to 70\% accuracy).

\item Predictive entropy (total uncertainty):
\begin{equation}
H_{\mathrm{pred}} = \frac{1}{N_{\mathrm{test}}}\sum_{i=1}^{N_{\mathrm{test}}} \Bigl( -\sum_{j\in\sY} \widehat{f}_\theta(j\mid x_i)\log\widehat{f}_\theta(j\mid x_i) \Bigr).
\end{equation}
This quantifies the total uncertainty (aleatoric + epistemic) in the predictive distribution. High entropy means the model is uncertain about the outcome.

\item Expected entropy (average aleatoric uncertainty):
\begin{equation}
H_{\mathrm{exp}} = \frac{1}{N_{\mathrm{test}}}\sum_{i=1}^{N_{\mathrm{test}}} \frac{1}{S}\sum_{\ell=1}^S \Bigl( -\sum_{j\in\sY} f_{\theta^{(\ell)}}(j\mid x_i)\log f_{\theta^{(\ell)}}(j\mid x_i) \Bigr).
\end{equation}
This averages the entropy of each individual sample's predictive distribution, capturing the average aleatoric uncertainty (inherent noise in the data). It is typically lower than the predictive entropy because the latter also includes the variation between samples.

\item Mutual information (epistemic uncertainty):
\begin{equation}
\mathrm{MI} = H_{\mathrm{pred}} - H_{\mathrm{exp}}.
\end{equation}
This difference between total uncertainty and average aleatoric uncertainty quantifies how much the model's uncertainty about the parameters (epistemic uncertainty) contributes to the predictive variance. Higher values indicate that the samples disagree more, i.e., the model is uncertain about the parameters themselves.

\item Epistemic standard deviation (lower values indicate less parameter-induced variability):
\begin{equation}
\mathrm{EpistemicStd} = \frac{1}{N_{\mathrm{test}}}\sum_{i=1}^{N_{\mathrm{test}}} \frac{1}{|\sY|}\sum_{j\in\sY} \sqrt{ \frac{1}{S} \sum_{\ell=1}^S \bigl( f_{\theta^{(\ell)}}(j\mid x_i) - \widehat{f}_\theta(j\mid x_i) \bigr)^2 }.
\end{equation}
This is a direct measure of parameter-induced variability in the predictions. Larger values indicate greater epistemic uncertainty.
\end{itemize}

\end{document}

%% file: tables/ex1_scores.tex
\begin{table}[htbp]
\centering
\footnotesize
\begin{tabular}{lrrrrrrrr}
\toprule
                    &       Accuracy &          Brier &            NLL &            ECE &   Pred entropy &    Exp entropy &    Mutual info &  Epistemic std\\
\midrule
NUTS                &         0.9304 &         0.0501 &         0.1672 &         0.0210 &         0.2149 &         0.2034 &         0.0115 &         0.0388\\
NUTS-mean           &         0.9304 &         0.0501 &         0.1658 &         0.0160 &         0.2006 &         0.2006 &         0.0000 &         0.0000\\
SMP-none            &         0.9294 &         0.0646 &         0.2320 &         0.0922 &         0.3602 &         0.1893 &         0.1709 &         0.1991\\
SMP-fixed-diag      &         0.9302 &         0.0508 &         0.1692 &         0.0219 &         0.2139 &         0.2008 &         0.0131 &         0.0392\\
SMP-fixed-block     &         0.9304 &         0.0510 &         0.1771 &         0.0343 &         0.2548 &         0.2177 &         0.0371 &         0.0846\\
SMP-fixed-dense     &         0.9194 &         0.0929 &         0.3386 &         0.1698 &         0.5317 &         0.1470 &         0.3847 &         0.3595\\
SMP-ema-diag        &         0.9300 &         0.0513 &         0.1719 &         0.0254 &         0.2217 &         0.2013 &         0.0203 &         0.0501\\
SMP-ema-block       &         0.9292 &         0.0524 &         0.1867 &         0.0451 &         0.2862 &         0.2403 &         0.0458 &         0.0973\\
SMP-ema-dense       &         0.9302 &         0.0501 &         0.1664 &         0.0183 &         0.2055 &         0.2043 &         0.0013 &         0.0117\\
SMP-periodic-diag   &         0.9296 &         0.0508 &         0.1695 &         0.0230 &         0.2152 &         0.2010 &         0.0142 &         0.0413\\
SMP-periodic-block  &         0.9298 &         0.0506 &         0.1719 &         0.0273 &         0.2342 &         0.2103 &         0.0239 &         0.0677\\
SMP-periodic-dense  &         0.9260 &         0.0645 &         0.2434 &         0.0925 &         0.3992 &         0.1762 &         0.2230 &         0.2555\\
\bottomrule
\end{tabular}
\caption{Predictive performance metrics across methods. Toy example.}
\label{tab:ex1_scores}
\end{table}

%% file: tables/ex1_variances.tex
\begin{table}[htbp]
\centering
\footnotesize
\begin{tabular}{lrr}
\toprule
                    &  Params variance &  Logits variance\\
\midrule
NUTS                &           0.3985 &           0.9526\\
NUTS-mean           &              NaN &              NaN\\
SMP-none            &           0.0022 &           9.4562\\
SMP-fixed-diag      &           0.0004 &           0.5015\\
SMP-fixed-block     &           0.1644 &           0.9253\\
SMP-fixed-dense     &           0.7334 &         128.8556\\
SMP-ema-diag        &           0.0007 &           0.8490\\
SMP-ema-block       &           0.1768 &           1.6986\\
SMP-ema-dense       &           0.0250 &           0.1335\\
SMP-periodic-diag   &           0.0004 &           0.5544\\
SMP-periodic-block  &           0.0961 &           8.9898\\
SMP-periodic-dense  &           0.2953 &          37.9504\\
\bottomrule
\end{tabular}
\caption{Mean parameter and logit variances across methods. Toy example.}
\label{tab:ex1_variances}
\end{table}

%% file: tables/ex1_elapsed.tex
\begin{table}[htbp]
\centering
\footnotesize
\begin{tabular}{lr}
\toprule
                    &  Elapsed time\\
\midrule
NUTS                &      405.0298\\
NUTS-mean           &        0.0000\\
SMP-none            &      107.3516\\
SMP-fixed-diag      &      145.8605\\
SMP-fixed-block     &      150.3850\\
SMP-fixed-dense     &      162.2861\\
SMP-ema-diag        &      149.0915\\
SMP-ema-block       &      163.3505\\
SMP-ema-dense       &      186.2400\\
SMP-periodic-diag   &      252.7232\\
SMP-periodic-block  &      260.3604\\
SMP-periodic-dense  &      286.5893\\
\bottomrule
\end{tabular}
\caption{Elapsed time in seconds across methods. Note: NUTS run with 1 thread, SMP with 6 threads. Toy example.}
\label{tab:ex1_elapsed}
\end{table}

%% file: tables/ex2_scores.tex
\begin{table}[htbp]
\centering
\footnotesize
\begin{tabular}{lrrrrrrrr}
\toprule
                   &       Accuracy &          Brier &            NLL &            ECE &   Pred entropy &    Exp entropy &    Mutual info &  Epistemic std\\
\midrule
MAP                &         0.9877 &         0.0189 &         0.0380 &         0.0012 &         0.0414 &         0.0414 &         0.0000 &         0.0000\\
SMP-none           &         0.9877 &         0.0190 &         0.0380 &         0.0011 &         0.0418 &         0.0416 &         0.0002 &         0.0004\\
SMP-fixed-diag     &         0.9837 &         0.4478 &         0.9907 &         0.6050 &         1.9379 &         0.0608 &         1.8770 &         0.2631\\
SMP-ema-diag       &         0.9797 &         0.1808 &         0.4582 &         0.3266 &         1.1462 &         0.2383 &         0.9079 &         0.1488\\
SMP-periodic-diag  &         0.9863 &         0.0312 &         0.0940 &         0.0581 &         0.2730 &         0.0994 &         0.1737 &         0.0410\\
\bottomrule
\end{tabular}
\caption{Predictive performance metrics across methods. MNIST example.}
\label{tab:ex2_scores}
\end{table}

%% file: tables/ex2_variances.tex
\begin{table}[htbp]
\centering
\footnotesize
\begin{tabular}{lrr}
\toprule
                   &  Params variance &  Logits variance\\
\midrule
MAP                &              NaN &              NaN\\
SMP-none           &         0.000000 &         0.008651\\
SMP-fixed-diag     &         0.000151 &       101.894676\\
SMP-ema-diag       &         0.000243 &        18.880239\\
SMP-periodic-diag  &         0.000025 &         2.776681\\
\bottomrule
\end{tabular}
\caption{Mean parameter and logit variances across methods. MNIST example.}
\label{tab:ex2_variances}
\end{table}

%% file: tables/ex2_elapsed.tex
\begin{table}[htbp]
\centering
\footnotesize
\begin{tabular}{lr}
\toprule
                   &  Elapsed time\\
\midrule
MAP                &          0.00\\
SMP-none           &       4130.18\\
SMP-fixed-diag     &       4619.78\\
SMP-ema-diag       &       6090.22\\
SMP-periodic-diag  &       6902.81\\
\bottomrule
\end{tabular}
\caption{Elapsed time in seconds across methods. MNIST example.}
\label{tab:ex2_elapsed}
\end{table}